\documentclass[12pt]{iopart}

\usepackage{pstricks,graphicx,bbm,amssymb, iopams}
\usepackage[utf8]{inputenc}
\usepackage{hyperref}
\usepackage[english]{babel}
\usepackage{color}
\usepackage{cite}

\def\Id{{\mathbbm 1}}
\def\dag{^{\dagger}}
\def\PGM{{\rm PGM}}
\def\HET{{\rm het}}
\def\opt{{\rm KOR}}
\def\disp{{\rm disp}}
\def\p{{\rm p}}
\def\Re{{\rm Re \, }}
\def\Im{{\rm Im \,}}

\begin{document}
\title[Optimizing state-discrimination receivers for continuous-variable$\ldots$]{Optimizing state-discrimination receivers for continuous-variable quantum key distribution over a wiretap channel}

\author{M.~N.~Notarnicola$^{1}$, M.~Jarzyna$^{2}$, S.~Olivares$^{1}$ and K.~Banaszek$^{2,3}$}
\address{$^{1}$Dipartimento di Fisica ``Aldo Pontremoli'',
Universit\`a degli Studi di Milano and INFN Sezione di Milano, via Celoria 16, I-20133 Milano, Italy}
\address{$^{2}$Centre for Quantum Optical Technologies, Centre of New Technologies, University of Warsaw, Banacha 2c, 02-097 Warszawa, Poland}
\address{$^{3}$Faculty of Physics, University of Warsaw, Pasteura 5, 02-093 Warszawa, Poland}
\ead{m.jarzyna@cent.uw.edu.pl}

\begin{abstract}
We address a continuous-variable quantum key distribution (CV-QKD) protocol employing quaternary phase-shift-keying (QPSK) of coherent states and a non-Gaussian measurement inspired by quantum receivers minimizing the error probability in a quantum-state-discrimination scenario. We consider a pure-loss quantum wiretap channel, in which a possible eavesdropper is limited to collect the sole channel losses. We perform a characterization of state-discrimination receivers and design an optimized receiver maximizing the asymptotic secure key rate (SKR),
namely the key-rate optimized receiver (KOR), comparing its performance with respect to the pretty good measurement (PGM) and the heterodyne-based protocol. We show that the KOR increases the 
SKR for metropolitan-network distances. Finally, we also investigate the implementations of feasible schemes, such as the displacement feed-forward receiver, obtaining an increase in the 
SKR in particular regimes.
\end{abstract}


\maketitle

\section{Introduction}
Quantum key distribution (QKD) is the art of sharing a secure key between two distant parties in the presence of an untrusted channel \cite{Gisin2002, Scarani2009, Pirandola2020, Xu2020}. In a QKD scenario, a transmitter (Alice) sends quantum states through a quantum channel to a receiver (Bob), who performs a suitable measurement to extract a set of correlated data. The security of this scheme against any potential attack by an eavesdropper (Eve) is guaranteed by the laws of quantum mechanics \cite{Wootters1982, Grosshans2001}.

Within the field, continuous-variable QKD (CV-QKD)\cite{Laudenbach2018} is of particular interest due to its feasibility with the technologies commonly employed in optical communications \cite{Braunstein2005}. Indeed, in CV-QKD information is encoded onto coherent states \cite{Olivares2021}, that is, laser pulses, just as in classical communication systems.
Several CV-QKD protocols have been proposed in literature, employing either Gaussian modulation of coherent states \cite{Grosshans2002, Grosshans2003-1, Grosshans2003-2, Weedbrook2004, Grosshans2005} or discrete modulation formats \cite{Leverrier2009, Sych2010, Leverrier2011, Becir2012, Ghorai2019, Lin2019, Liao2020, Denys2021, Upadhyaya2021, Liu2021, Lupo2022, Notarnicola2022}. Nevertheless, all these schemes assume homodyne or heterodyne (equivalent to double homodyne) detection at Bob's side, and unconditional security proofs are currently established only for Gaussian measurements \cite{Navascues2006, LeverrierThesis, GarciaPatron2006}.

On the other hand, in many other frameworks non-Gaussian measurements often outperform Gaussian ones, and this makes it interesting to investigate their role also for CV-QKD. A relevant example is provided by quantum state discrimination theory, in which Alice encodes information onto non-orthogonal states and the task is to design an efficient receiver minimizing the error probability of Bob's decision \cite{Holevo1973, Yuen1975, Helstrom1976, Bergou2010, Cariolaro2015}. This optimal receiver has been shown to be non-Gaussian, whereas Gaussian receivers are limited by the so-called shot noise limit (SNL) \cite{Cariolaro2015}. However, deriving the optimum measurement is rather cumbersome as it requires convex semidefinite programming \cite{Eldar2003, Assalini2010,Cariolaro2015}. As a consequence, there have also been developed suboptimal methods such as the pretty good measurement (PGM) method, which provides an upper bound to the minimum error probability and still beats the SNL \cite{Eldar2001, Eldar2002, Cariolaro2015}. 
Finally, feasible receivers have been proposed for both binary \cite{Kennedy1973, DiMario2018-1, Notarnicola2023, NotarnicolaFF} and quaternary \cite{Izumi2012, Becerra2013, DiMario2018-2, Izumi2020} coherent-state discrimination, exploiting linear optics and photodetection, thus being easier to implement into practice.
Beside quantum communications, in the existing literature state-discrimination receivers have also been investigated for CV-QKD \cite{Liao2018}.

In this paper, we propose a new quantum state-discrimination receiver for CV-QKD, namely, the key-rate optimized receiver (KOR). In more detail, we consider a CV-QKD protocol employing a quaternary phase-shift-keying (QPSK) modulation of coherent states \cite{Leverrier2009, Ghorai2019, Lin2019, Liao2020, Denys2021, Upadhyaya2021, Lupo2022} in which Bob implements the KOR rather than a Gaussian measurement. We compute the 
secure key rate (SKR), that is, the asymptotic length of the secret key per unit time slot, assuming a quantum wiretap channel \cite{Cai2004, Holevo2012, Pan2020}, which represents a realistic paradigm for many different situations, e.g. free-space optical communications \cite{Banaszek2021}. 
For the sake of simplicity, we consider a pure-loss channel, in order to deal with discrimination of pure states at Bob's side. This latter assumption depicts a simplified scenario, providing a cornerstone fostering more advanced developments.
We design the KOR to maximize the 
SKR of the addressed protocol and compare it with the PGM, showing that both these non-Gaussian measurements improve the 
SKR in the metropolitan-network distance regime with respect to the heterodyne-based protocol. We also investigate the performance of some feasible quantum receivers, by considering the feed-forward displacement receiver proposed in \cite{Izumi2012}.

The structure of the paper is the following. First, in Section~\ref{sec:QDT} we briefly outline the general structure of quantum state-discrimination receivers that will be employed throughout the work. Thereafter, Section~\ref{sec:Proto} presents the scheme of the protocol under investigation and describes the construction of the KOR. Then, in Section~\ref{sec:Results} we compare the performance of the designed receiver with the same protocol employing heterodyne detection, whereas in Section~\ref{sec:Izumi} we compute the 
SKR
for the feed-forward displacement receiver. Finally, in Section~\ref{sec:Conc} we summarize the results and draw some conclusions and outlooks.

\section{General structure of a state-discrimination receiver}\label{sec:QDT}
In this paper, we investigate the potentiality of state-discrimination receivers for secure quantum communications and propose a new quantum receiver, the KOR, for CV-QKD protocols employing discrete modulation.
To begin with, in this section, we present the general structure of a state-discrimination receiver within the framework of quantum state discrimination theory, or quantum decision theory \cite{Holevo1973, Yuen1975, Eldar2001, Eldar2002, Eldar2003, Assalini2010, Cariolaro2015}.
 
Generally speaking, we have a source emitting $M$ non-orthogonal linearly independent quantum states $\{ |\gamma_k\rangle \}_k$, $k=0,\ldots,M-1$, with a priori probabilities $q_k$, $0\le q_k\le 1$, and the task is to design a quantum receiver. That is, we look for a positive-operator-valued measure (POVM) $\{\Pi_j\}_j$, $j=0,\ldots,M-1$, $\Pi_j\ge 0$ and $\sum_j \Pi_j = \hat{\Id}$, $\hat{\Id}$ being the identity operator over the whole Hilbert space, such that registering the outcome $j$ infers the state generated by the source to be $|\gamma_j\rangle$.
When the outcome $j$ is obtained, we underline that the receiver makes the decision $j$ even if the state $k\neq j$ was sent, resulting in a decision error.

For the decision between pure states, the Kennedy theorem \cite{Kennedy1973, Cariolaro2015} proves an optimum POVM to be 1-~rank, namely, $\Pi_j=|\mu_j\rangle\langle \mu_j|$, expressed in terms of a set of measurement vectors $\{|\mu_j\rangle\}_j$.
As a consequence, the problem may be recast into a geometric optimization task. Indeed, we introduce the state and measurement (row) matrices
\begin{eqnarray}
\Gamma = \bigg(|\gamma_0\rangle, \ldots, |\gamma_{M-1}\rangle\bigg) \quad \mbox{and} \quad \mathbb{M} = \bigg(|\mu_0\rangle, \ldots, |\mu_{M-1}\rangle\bigg) \, ,
\end{eqnarray}
respectively. Moreover, it is not restrictive to reduce the problem to the $M$ dimensional subspace spanned by the encoded states, ${\cal S} = {\rm span} \{ |\gamma_k\rangle : k=0,\ldots, M-1 \}$ \cite{Cariolaro2011, Cariolaro2015}. Accordingly, the measurement vectors are expressible as a linear combination of the state vectors, $|\mu_j\rangle = \sum_k a_{kj} |\gamma_k\rangle$, $a_{kj} \in \mathbb{C}$, or equivalently,
\begin{eqnarray}\label{eq:Mmatrix}
\mathbb{M}= \Gamma \, A \, ,
\end{eqnarray}
with $A$ being an $M\times M$ matrix with coefficients $(a_{kj})_{k,j}$. The identity resolution constraint now becomes $\sum_j \Pi_j= \mathbb{P}_{\cal S}$, $ \mathbb{P}_{\cal S}$ being the projection operator onto subspace $\cal S$.

In turn, the conditional probability of obtaining outcome $j$ if the $k$-th state is probed is given by
\begin{eqnarray}\label{eq:CondP}
p(j|k)= \Tr \bigg[ |\gamma_k\rangle \langle \gamma_k| \, \Pi_j \bigg]= |B_{kj}|^2 \, ,
\end{eqnarray}
where
\begin{eqnarray}
B= \mathbb{M}\dag \, \Gamma = A\dag \, G \, ,
\end{eqnarray}
$G$ being the $M\times M$ Gram matrix:
\begin{eqnarray}\label{eq:Gmatrix}
G= \Gamma\dag \, \Gamma = \bigg(\langle \gamma_l | \gamma_k\rangle\bigg)_{l,k=0,\ldots,M-1} \, ,
\end{eqnarray}
that is the matrix of the overlaps between the encoded states. Thus, a quantum receiver is completely characterized by its corresponding matrix $A$, subject to the constraint
\begin{eqnarray}\label{eq:ConstA}
A A\dag = G^{-1} \, ,
\end{eqnarray}
guaranteeing the identity resolution of the resulting POVM, as derived in~\ref{app:AdagA}.

The scenario may be further simplified if the constellation of states $\{|\gamma_k\rangle\}_k$ exhibits the geometric uniform symmetry (GUS) \cite{Cariolaro2015}, namely if there exists a symmetry operator $S$ such that
\begin{eqnarray}
|\gamma_k\rangle= S^k \, |\gamma_0\rangle \quad \mbox{and} \quad S^M=\hat{\Id} \, , \qquad (k=0,\ldots,M-1) \, .
\end{eqnarray}
In this case, we may safely assume also the measurement vectors to exhibit the GUS for the same operator $S$, thus the POVM $\{\Pi_j\}_j$ is identified by a single ``reference" measurement vector
\begin{eqnarray}
|\mu_0\rangle = \sum_{k=0}^{M-1} a_{k0} \, |\gamma_k\rangle \, ,
\end{eqnarray}
$a_{k0} \in \mathbb{C}$, while all the others will be retrieved as $|\mu_j\rangle= S^j |\mu_0\rangle$, $j=0,\ldots, M-1$. Consequently, $A$ is a circulant matrix, having the form
\begin{eqnarray}
A= 
\left(
\begin{array}{cccc} 
a_{00} & a_{M-1 \,0} & \ldots  & a_{10} \\
a_{10} & a_{00} &   \ldots & a_{20} \\
\vdots & \vdots & \ddots & \vdots \\
a_{M-1 \, 0} & a_{M-2 \,0} & \ldots & a_{00} \\
\end{array}
\right) \, ,
\end{eqnarray}
which is diagonalizable by the unitary matrix $U=  \mathbb{F}^{-1}$, $\mathbb{F}=(\rme^{-\rmi 2\pi j k/M}/\sqrt{M})_{jk}$, $j,k=0,\ldots,M-1$, being the discrete Fourier transform matrix \cite{Davis1970}. 
Its eigendecomposition is given by $A= U \Lambda U\dag$, where $\Lambda= {\rm diag}(\lambda_0,\ldots, \lambda_{M-1})$, is the diagonal matrix composed of the eigenvalues $\{\lambda_j\}_j$ of $A$. Furthermore, circulant matrices form a commutative algebra \cite{Davis1970}, thus $A\dag$ is also circulant and $[A,A\dag]=0$.
Thereby, $A\dag= U \Lambda\dag U\dag$ and Equation~(\ref{eq:ConstA}) becomes:
\begin{eqnarray}
U \, |\Lambda|^2 \, U\dag = G^{-1} \, ,
\end{eqnarray}
where $|\Lambda|^2= {\rm diag}(|\lambda_0|^2,\ldots, |\lambda_{M-1}|^2)$. We conclude that $A$ and $G^{-1}$ are simultaneously diagonalizable and $|\lambda_j|^2= g_j^{-1}$, $\{g_j\}_j$ being the eigenvalues of the Gram matrix~(\ref{eq:Gmatrix}) listed in increasing order, that is, $g_0\ge g_1\ge \ldots\ge g_{M-1}$. In conclusion, the matrix $A$ may be re-expressed in the following form:
\begin{eqnarray}\label{eq:Aphi}
A\equiv A_{\boldsymbol{\phi}}=U \, \Lambda_{\boldsymbol{\phi}} \, U\dag \,,
\end{eqnarray}
where 
\begin{eqnarray}
\Lambda_{\boldsymbol{\phi}}= {\rm diag}\Bigg(\bigg\{\lambda_j^{(\boldsymbol{\phi})} \bigg\}_{j=0,\ldots, M-1}\Bigg) \, ,
\end{eqnarray}
and
\begin{eqnarray}
\lambda_j^{(\boldsymbol{\phi})} = \rme^{\rmi \phi_j} g_j^{-1/2} \, ,
\end{eqnarray}
in which the relative phases $\boldsymbol{\phi}=(\phi_0,\ldots,\phi_{M-1})$ provide the only free parameters. Furthermore, the matrix $A_{\boldsymbol{\phi}}$ is defined up to an overall phase due to~(\ref{eq:Mmatrix}), therefore we may fix $\phi_0=0$, ending up with $M-1$ phases whose value can be arbitrarily chosen.

We conclude that, in the presence of GUS, every quantum receiver is ultimately identified by the set of phases $\boldsymbol{\phi}$, which may be properly chosen to optimize a desired figure of merit, according to the context under investigation. In the existing literature, the typical figure of merit of quantum discrimination theory is the error probability
\begin{eqnarray}\label{eq:ErrorP}
P_{\rm err} = 1 - \sum_{k=0}^{M-1} q_k \, p(k|k)\, ,
\end{eqnarray}
representing the probability of inferring a symbol $j\neq k$ if the $k$-th state is sent. In this case, the optimal matrix $A$ minimizing~(\ref{eq:ErrorP}) may be retrieved via convex semidefinite programming \cite{Cariolaro2015, Eldar2003, Assalini2010}. Alternatively, simpler suboptimal schemes have been proposed, providing an upper bound of the minimum error probability. The most relevant is the PGM method \cite{Cariolaro2015, Eldar2001,Eldar2002}, leading to a suboptimal POVM which coincides with the optimal measurement for constellations exhibiting GUS \cite{Cariolaro2015}. The matrix $A$ for the PGM reads:
\begin{eqnarray}\label{eq:APGM}
A_\PGM = G^{-1/2} \, ,
\end{eqnarray}
which may be retrieved from~(\ref{eq:Aphi}) by setting all phases to zero, $\boldsymbol{\phi}=(0,\ldots,0)$.

In the following section, we discuss a different approach and adopt a quantum receiver described by~(\ref{eq:Aphi}) to maximize the 
SKR of the four-state CV-QKD protocol, optimizing the free phases to design the optimized discrete receiver for CV-QKD, namely, the KOR.

\section{CV-QKD with state-discrimination receivers over a wiretap channel}\label{sec:Proto}
\begin{figure}
\begin{center}
\includegraphics[width=0.8\textwidth]{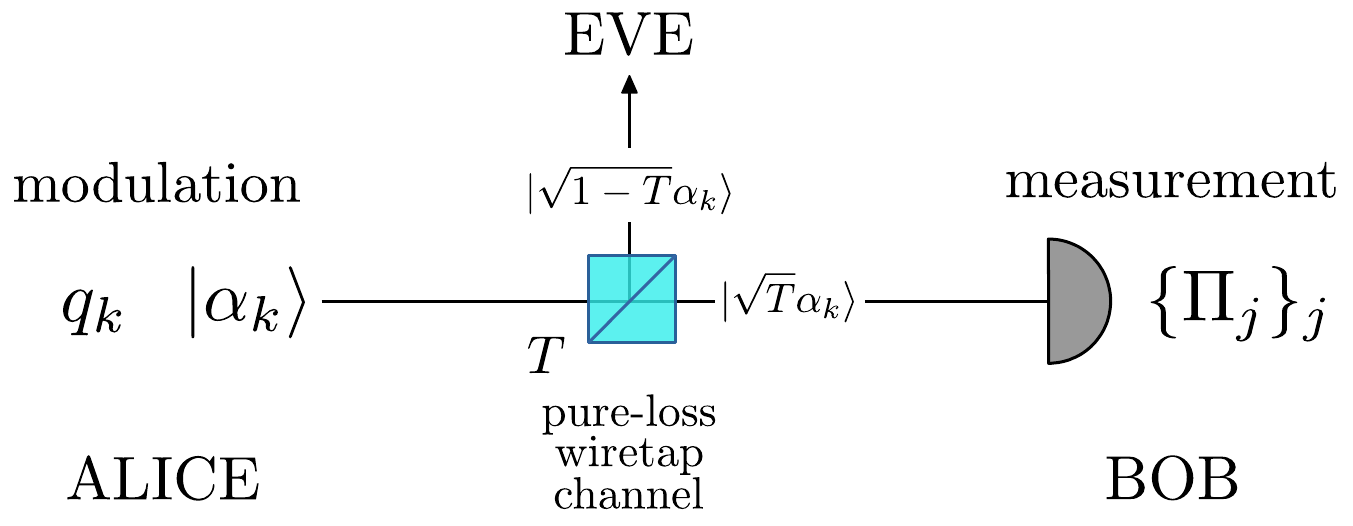}
\end{center}
\caption{Scheme of the CV-QKD protocol discussed in the paper. Alice generates one of the coherent states $|\alpha_k\rangle$, $k=0,\ldots, M-1$, with uniform probability $q_k=1/M$ and sends it via the quantum wiretap channel to Bob, who performs the POVM $\{\Pi_j\}_j$.}\label{fig:01-Protocol}
\end{figure}

As mentioned in the previous section, we design the KOR to optimize the 
SKR of the four-state protocol depicted in Figure~\ref{fig:01-Protocol}, in which the sender, Alice, employs the QPSK modulation. The latter is a special case of phase-shift keying (PSK) modulation in which information is encoded in one of $M$ coherent states
\begin{eqnarray}
|\alpha_k\rangle= |\alpha \, \rme^{\rmi \pi (2k+1)/M} \rangle \, , \qquad (k=0,\ldots, M-1) \, , 
\end{eqnarray}
where $\alpha\ge0$, generated with equal a priori probabilities $q_k=1/M$. That is, the PSK constellation is composed of $M$ coherent states with the same energy $\alpha^2$ and phase-shifted by $\theta=2\pi/M$; therefore, it satisfies the GUS for the phase-shift symmetry operator $S_\theta= \exp(- \rmi \, \theta \, \hat{n})$, $\hat{n}$ being the photon-number quantum operator \cite{Cariolaro2015}. The QPSK is a special case of PSK with $M=4$.

After the modulation stage, Alice injects the signals into an untrusted quantum channel and the receiver, Bob, probes the received pulses via a state-discrimination receiver, namely, a finite-valued POVM $\{\Pi_j\}_j$ with the properties described in Section~\ref{sec:QDT}.
However, already existing protocols \cite{Grosshans2002, Grosshans2003-1, Grosshans2003-2, Weedbrook2004, Grosshans2005, Leverrier2009, Leverrier2011, Becir2012, Ghorai2019, Denys2021} usually assume Gaussian POVMs at the receiver, either homodyne or heterodyne detection, and unconditional security proofs are guaranteed only for Gaussian measurements \cite{Navascues2006, LeverrierThesis, GarciaPatron2006}. 
In fact, under the unconditional security paradigm, one considers the most powerful eavesdropping attack, where an eavesdropper, Eve, performs any arbitrary channel manipulation that preserves the local statistics at Alice’s and Bob’s sides. Physical layer security is then addressed via a constrained functional optimization over all possible channels, leading to the Devetak-Winter (DW) bound \cite{Devetak2005}. This task has been obtained for protocols employing Gaussian measurements under the assumption of either a linear \cite{Leverrier2009} or nonlinear channel \cite{Ghorai2019}, thanks to the optimality of Gaussian attacks \cite{Navascues2006, LeverrierThesis, GarciaPatron2006}. Recently, a tight lower bound to the DW bound has also been obtained via nonlinear convex optimization algorithms and without invoking Gaussian optimality \cite{Lin2019}.

Nevertheless, in practical scenarios it is justified to address physical layer security under a wider set of assumptions, including restricted eavesdropping \cite{Pan2020} and different trust levels of the infrastructure \cite{Pirandola2021}, computing the SKR under the so-called composable security paradigm. For instance, in a typical satellite QKD attack, Eve intercepts a part of the signal not being captured by Bob’s telescope \cite{Pirandola2021-Free}. Similarly, in the fiber based protocols we may assume an attack to be performed without breaking the structure of the fiber to implement arbitrary signal manipulation. 
In the recent literature, secure key distillation in these practical conditions is gaining much interest, with particular reference to wiretap channels \cite{Pan2020} modeling the previous realistic assumptions, and passive eavesdropping \cite{Derkach2021, Kundu2023}, which have also been tested experimentally \cite{Fujiwara2018}.

Given the present scenario, CV-QKD with non-Gaussian measurements still remains an open problem since the Gaussian optimality theorem does not hold anymore, and the DW bound can only be directly evaluated with the advanced methods presented in \cite{Lin2019}.
In light of this, here we start from scratch and address composable security, by considering a quantum wiretap channel \cite{Cai2004, Holevo2012, Pan2020}. In this scenario, we consider a restricted eavesdropping strategy where Eve may only collect the lost fraction of the encoded signals without performing arbitrary channel manipulation. Once again, we remark that this represents a realistic scenario which is worth of interest for practical conditions, but does not guarantee unconditional security. 
In turn, the purpose of our research is to show that, in particular conditions, it is possible to theoretically design a suitable measurement outperforming the conventional quadrature detection schemes.

Moreover, we also perform a further assumption and consider a pure-loss channel, described as a beam splitter with transmissivity
\begin{eqnarray}
T= 10^{-\kappa d/10} \,,
\end{eqnarray}
$d$ being the transmission distance (expressed in km) and $\kappa=0.2$ dB/km is the loss rate of common fibers at telecom wavelength \cite{Lodewyck2005,Lodewyck2007, Banaszek2020, Agrawal2002, Jouguet2013}.
This case is worth of investigation for a twofold reason. At first, in the recent literature there has been a revived interest in passive eavesdropping strategies \cite{Banaszek2021, Derkach2021, Kundu2023, Medlock2021, Ghalaii2022, Jarzyna2023}. Secondly, passive eavesdropping can be used as the first stepping stone to identify scenarios where a potential advantage of quantum receivers may be substantial, even in the possible presence of nonzero excess noise.

Provided these two assumptions, in the following we construct the optimized POVM that describes the KOR, and show it to bring advantages in some particular regimes. 

In the presence of a pure-loss quantum wiretap channel, if Alice generates the state $|\alpha_k\rangle$, the transmitted fraction $|\alpha_k^{(t)}\rangle = |\sqrt{T} \alpha_k\rangle$ reaches Bob, whereas Eve receives the reflected part $|\alpha_k^{(r)}\rangle = |\sqrt{1-T} \alpha_k\rangle$, as shown in Figure~\ref{fig:01-Protocol}.
Accordingly, the dimensional reduction holds \cite{Cariolaro2015,Cariolaro2011}, and we restrict our attention to the $M$-dimensional subspace spanned by the transmitted pulses, namely, ${\cal S} = {\rm span} \{ |\alpha_k^{(t)}\rangle : k=0,\ldots, M-1 \}$ and apply the results of Section~\ref{sec:QDT}.

To form a truly identity-resolving set, the POVM elements $\{\Pi_j\}_j$, $j=~0,\ldots, M-1$, shall be complemented with an $(M+1)$-th inconclusive element $\Pi_M= \hat{\Id}- \mathbb{P}_{\cal S}$, $\mathbb{P}_{\cal S}$ being the projection operator onto $\cal S$. For the case under investigation, $\Pi_M$ is irrelevant, and we will neglect it in the following. However, this would not be true anymore in the presence of a channel excess noise; thus, registering a zero probability for this additional outcome may provide a useful way to check the reasonableness of the pure-loss hypothesis in a realistic implementation of the proposed protocol.

\subsection{Construction of the KOR}\label{sec:POVM}
In our protocol Bob shall employ an optimized POVM to perform discrimination among the transmitted pulses, described by the state vector $\Gamma = (|\alpha_0^{(t)}\rangle, \ldots, |\alpha_{M-1}^{(t)}\rangle)$ and the Gram matrix:
\begin{eqnarray}\label{eq:GramT}
G=\Big(\big\langle \alpha^{(t)}_l \big| \alpha^{(t)}_k \big\rangle \Big)_{l,k=0,\ldots,M-1} \, ,
\end{eqnarray}
in which the overlap $G_{lk}=\langle \alpha^{(t)}_l \big| \alpha^{(t)}_k \big\rangle$ reads \cite{Serafini2017}:
\begin{eqnarray}
G_{lk}&= \exp\left\{-\frac12 \left|\alpha^{(t)}_k -\alpha^{(t)}_l \right|^2 + \frac12 \left[ \alpha^{(t)}_k  \left(\alpha^{(t)}_l\right)^*-\left(\alpha^{(t)}_k\right)^* \alpha^{(t)}_l \right] \right\} \nonumber \\[1.5ex]
&= \exp \left(-T \alpha^2 \left\{1-\cos\left[\frac{2\pi}{M}(k-l)\right] \right\} + \rmi \, T \alpha^2 \sin\left[\frac{2\pi}{M}(k-l)\right] \right)\, .
\end{eqnarray}

The constellation of transmitted pulses maintains the GUS for the phase-shift operator $S_\theta$, thus the set of measurement vectors $\mathbb{M}=~\{|\mu_j\rangle\}_j$, $j=0,\ldots,M-1$, also satisfies the GUS, and the corresponding matrix $A$ is in the form~(\ref{eq:Aphi}), according to the results of Section~\ref{sec:QDT}.
The KOR is obtained by optimizing the phases $\boldsymbol{\phi}$ to maximize the 
SKR, and is described by the optimized ``reference" measurement vector:
\begin{eqnarray}\label{eq:mu0phi}
|\mu_0^{(\boldsymbol{\phi})}\rangle &= \sum_{k=0}^{M-1} \left(A_{\boldsymbol{\phi}}\right)_{k0} \, \left|\alpha_k^{(t)}\right\rangle \\
&= \rme^{-T \alpha^2/2} \sum_{n=0}^{\infty} \frac{\left(\sqrt{T}\alpha_0\right)^n}{\sqrt{n!}} \,  \lambda_{(n-1) \bmod M}^{(\boldsymbol{\phi})} \, |n\rangle\, ,
\end{eqnarray}
where $a \bmod b$ is the modulo operation, returning the reminder of the division $a/b$, $a,b\in \mathbb{Z}$, and $\{|n\rangle\}_n$ is the photon-number basis.

Given the previous considerations, we compute the 
SKR of the discussed protocol, considering a reverse reconciliation scenario, which guarantees higher security in many existing protocols \cite{Grosshans2002, Grosshans2003-1, Grosshans2003-2}. 
Moreover, for the sake of simplicity, we perform the asymptotic key-rate calculation, where the channel parameters are known with no uncertainty. Under this paradigm, for a generic state-discrimination receiver described by the phase vector $\boldsymbol{\phi}$, the SKR reads:
\begin{eqnarray}\label{eq:KGR}
K(\boldsymbol{\phi},\alpha^2)= \beta I_{AB}(\boldsymbol{\phi},\alpha^2)- \chi_{BE}(\boldsymbol{\phi},\alpha^2)\, ,
\end{eqnarray}
where $I_{AB}$ and $\chi_{BE}$ are the mutual information between Alice and Bob and the Holevo information \cite{Holevo1998} between Bob and Eve, respectively, and $\beta\le 1$ is the reconciliation efficiency \cite{Leverrier2009, Leverrier2011}.

The mutual information reads:
\begin{eqnarray}\label{eq:IAB}
I_{AB}(\boldsymbol{\phi},\alpha^2)= H\left[p_B^{(\boldsymbol{\phi})}(j)\right] - \frac{1}{M} \sum_{k=0}^{M-1} H\left[p_{B|\alpha_k}^{(\boldsymbol{\phi})}(j)\right] \, ,
\end{eqnarray}
where
\begin{eqnarray}\label{eq:pB|A}
p_{B|\alpha_k}^{(\boldsymbol{\phi})}(j) 
= \left\langle \sqrt{T} \alpha_k \left|  \Pi_j \right| \sqrt{T} \alpha_k \right\rangle 
= \left|\left(A_{\boldsymbol{\phi}}\dag G\right)_{kj}\right|^2 \, ,
\end{eqnarray}
$G$ being the Gram matrix~(\ref{eq:GramT}), and
\begin{eqnarray}\label{eq:pB}
p_B^{(\boldsymbol{\phi})}(j) =\frac{1}{M} \sum_{k=0}^{M-1} p_{B|\alpha_k}^{(\boldsymbol{\phi})}(j) \, ,
\end{eqnarray}
are the conditional and overall probabilities of Bob's detection associated with outcome $j=0,\ldots,M-1$, respectively, and $H[p(x)]= -\sum_x p(x) \log_2 p(x)$ is the Shannon entropy of the probability distribution $p(x)$.

To compute the Holevo information shared between Bob and Eve, that is the maximum amount of information accessible to Eve, we approach the problem in the prepare-\&-measure picture \cite{Laudenbach2018, Grosshans2002, Grosshans2003-1} and obtain:
\begin{eqnarray}\label{eq:chiBE}
\chi_{BE}(\boldsymbol{\phi},\alpha^2)= S\left[\rho_E \right] - \sum_{j=0}^{M-1}  p_B^{(\boldsymbol{\phi})}(j) \, S\left[\rho_{E|j}^{(\boldsymbol{\phi})}\right] \, ,
\end{eqnarray}
where $\rho_{E|j}^{(\boldsymbol{\phi})}$ and $\rho_E$ are the conditional and overall Eve’s state, respectively, $p_B^{(\boldsymbol{\phi})}(j)$ is Bob’s probability distribution~(\ref{eq:pB}) and $S[\rho]= - \Tr[\rho \log_2 \rho]$ represents the von Neumann entropy associated with state $\rho$.
These two states may be retrieved from the joint state of Bob and Eve after the channel, that is:
\begin{eqnarray}
\rho_{BE}&=U_{BS}(T) \, \rho_A \otimes |0\rangle \langle 0 | \, U_{BS}(T)\dag \nonumber \\[1ex]
&= \frac{1}{M} \sum_{k=0}^{M-1} \left|\alpha_k^{(t)} \right\rangle\left \langle \alpha_k^{(t)} \right| \otimes  \left|\alpha_k^{(r)} \right\rangle \left\langle \alpha_k^{(r)} \right| \, ,
\end{eqnarray}
where $\rho_A= \sum_k q_k|\alpha_k\rangle \langle \alpha_k|$ is Alice's overall state, $|0\rangle$ is the vacuum state and $U_{BS}(T)$ is unitary operator associated with a beam splitter with transmissivity $T$ \cite{Olivares2021}, as displayed in Figure~\ref{fig:01-Protocol}. In turn, we have:
\begin{eqnarray}\label{eq:rhoE}
\rho_E= \Tr_B\left[\rho_{BE}\right] = \frac{1}{M} \sum_{k=0}^{M-1} \left|\sqrt{1-T} \alpha_k \right\rangle \left\langle \sqrt{1-T} \alpha_k \right| \, ,
\end{eqnarray}
and
\begin{eqnarray}\label{eq:rhoEcondB}
\rho_{E|j}^{(\boldsymbol{\phi})}&= \frac{1}{p_B^{(\boldsymbol{\phi})}(j)} \Tr_B\left[\rho_{BE} \, \Pi_j \otimes \hat{\Id}_E\right] \nonumber \\[1ex]
&= \frac{1}{M p_B^{(\boldsymbol{\phi})}(j)} \sum_{k=0}^{M-1} \, p_{B|\alpha_k}^{(\boldsymbol{\phi})}(j) \, \left|\sqrt{1-T} \alpha_k \right\rangle \left\langle \sqrt{1-T} \alpha_k \right| \, ,
\end{eqnarray}
$\Tr_B$ being the partial trace over Bob's mode and $\hat{\Id}_E$ being the identity operator over Eve's mode.
Finally, the von Neumann entropy of states~(\ref{eq:rhoE}) and~(\ref{eq:rhoEcondB}) may be computed with the methods outlined in \ref{app:Entropies}.

In our analysis, we are interested in the maximum achievable 
SKR as a function of the transmission distance $d$, therefore, in the end we will perform optimization over the free parameters, namely the phases $\boldsymbol{\phi}$ and the constellation energy $\alpha^2$. The final, optimized, 
SKR is therefore equal to
\begin{eqnarray}\label{eq:Kopt}
K_\opt= \max_{\boldsymbol{\phi},\, \alpha^2} \, K(\boldsymbol{\phi},\, \alpha^2) \, ,
\end{eqnarray}
with the optimized phases and modulation energy denoted by $\boldsymbol{\phi}_\opt= (0, \phi_1^{(\opt)},\ldots, \phi_{M-1}^{(\opt)})$ and $\alpha^2_\opt$. 
As a consequence, we define the KOR via Equation~(\ref{eq:Aphi}), as the quantum receiver associated with the optimized phase vector $\boldsymbol{\phi}_\opt$. Furthermore, we compare the performance of the KOR with that associated with the PGM, for which the optimized 
SKR reads:
\begin{eqnarray}\label{eq:KPGM}
K_\PGM= \max_{\alpha^2} \,  K(\boldsymbol{\phi=0},\, \alpha^2) \, ,
\end{eqnarray}
with zero phases and the optimized energy $\alpha^2_\PGM$. The results obtained for the above two receivers are discussed in the following section, where we compare both 
SKRs with the key rate of the heterodyne-based protocol in order to highlight the advantages brought by the two non-Gaussian measurements.

The heterodyne-based protocol is analogous to the one discussed above and employs double homodyne detection at Bob's side, that is, a measurement of both field quadratures $q$ and $p$, retrieving a pair of real outcomes ${\bf x}=(x_B,y_B)\in\mathbb{R}^2$. 
We underline that, while both the KOR and the PGM are described in terms of a finite-valued POVM with $M$ possible outcomes, in the presence of heterodyne detection we have a continuous-variable measurement.
Therefore, in this case Bob's conditional probability reads:
\begin{eqnarray}
p_{B|\alpha_k}^{(\HET)}({\bf x}) 
= \frac{1}{4\pi \sigma_0^2} 
& \exp\left\{-\left[x_B-2\sigma_0 \sqrt{T}\,  \Re(\alpha_k)\right]^2/(4\sigma_0^2) \right\} \times \nonumber \\[.5ex]
& \exp\left\{-\left[y_B-2\sigma_0 \sqrt{T} \, \Im(\alpha_k)\right]^2/(4\sigma_0^2)\right\} \, ,
\end{eqnarray}
$\sigma_0^2$ being the shot-noise variance \cite{Olivares2021}, which from now on will be taken equal to $1$, meaning, we perform calculations in shot-noise units (SNU).
The obtained mutual information is, similarly as in (\ref{eq:IAB}), given by:
\begin{eqnarray}
I_{AB}^{(\HET)} (\alpha^2)= H\left[p_B^{(\HET)} ({\bf x})\right] - \frac{1}{M} \sum_{k=0}^{M-1} H\left[p_{B|\alpha_k}^{(\HET)} ({\bf x})\right] \, ,
\end{eqnarray}
with $p_B^{(\HET)} ({\bf x})= M^{-1} \sum_k p_{B|\alpha_k}^{(\HET)}({\bf x})$. Instead, the Holevo information becomes:
\begin{eqnarray}\label{eq:holevo_het}
\chi_{BE}(\alpha^2)= S\left[\rho_E \right] - \int_{\mathbb{R}^2} d^2{\bf x} \,\, p_{B|\alpha_k}^{(\HET)}({\bf x}) \,\, S\left[\rho_{E|{\bf x}}\right] \, ,
\end{eqnarray}
with $\rho_E$ given in~(\ref{eq:rhoE}) and
\begin{eqnarray}
\rho_{E|{\bf x}}
=& \frac{1}{M p_B^{(\HET)}({\bf x})} \,\, \sum_{k=0}^{M-1} p_{B|\alpha_k}^{(\HET)}({\bf x}) \,\, \left|\sqrt{1-T} \alpha_k \right\rangle \left\langle \sqrt{1-T} \alpha_k \right| \, .
\end{eqnarray}
The integration in (\ref{eq:holevo_het}) can been performed numerically by exploiting the Simpson's rule \cite{Press2007}.
Finally, the resulting 
SKR is obtained as
\begin{eqnarray}\label{eq:Khet}
K_\HET= \max_{\alpha^2} \, \left[\beta I_{AB}^{(\HET)}(\alpha^2) -\chi_{BE}^{(\HET)}(\alpha^2) \right]\, ,
\end{eqnarray}
with the optimized modulation energy $\alpha^2_\HET$.

As a final remark, towards a realistic implementation of the present protocol, we underline that both the KOR and the PGM may not represent appropriate POVMs for the channel evaluation stage. Indeed, assuming that the channel properties do not change, Alice and Bob must estimate the channel parameters, which in the present case is limited to the sole transmissivity $T$. However, unlike homodyne and heterodyne detection, in principle the designed POVM $\{\Pi_j\}_j$ does not guarantee full channel characterization. This problem may be circumvented, at least for the asymptotic key rate calculation, 
by performing Gaussian detection on a small fraction of the exchanged pulses and reserving it for the channel estimation stage, whilst exploiting the non-Gaussian receiver only for the key extraction.
On the contrary, in the presence of a finite-size scenario, Alice and Bob estimate the channel transmissivity $T$ with a finite uncertainty $\Delta T$, thus leaving more space for Eve's intervention. Therefore, they employ a conservative strategy and compute the SKR by considering a lower value of the transmissivity, namely $T-\Delta T$. However, the main effect of this lower effective transmissivity is to reduce the range of distances for which the state-discrimination receivers outperforms the heterodyne protocol. Furthermore, the dataset for the key extraction is also finite, resulting in a lower SKR with respect to the asymptotic case.

\section{Results}\label{sec:Results}
\begin{figure}
\begin{center}
\includegraphics[width=\textwidth]{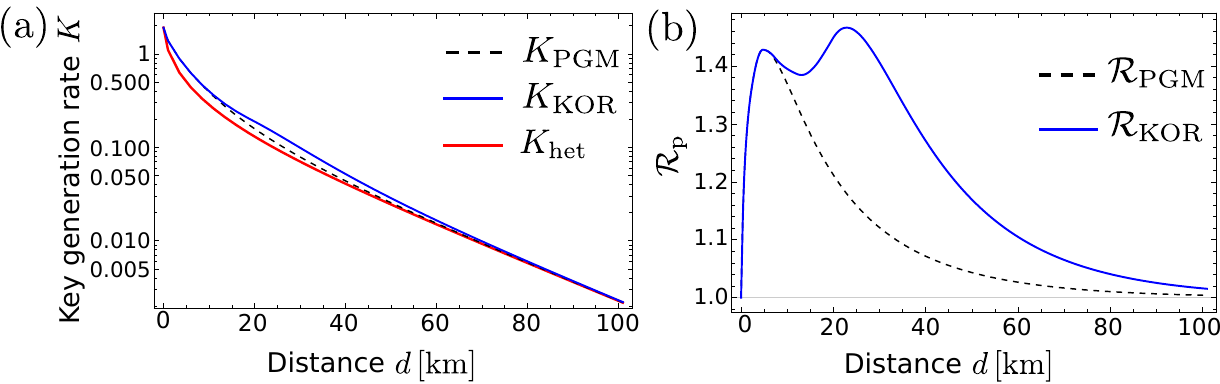}
\end{center}
\caption{(a) Logarithmic plot of $K_{\p}$, $\p=\PGM,\opt$, compared to $K_{\HET}$, as a function of the transmission distance $d$ in km. (b) Plot of the ratio ${\cal R}_{\p}$, $\p=\PGM,\opt$, as a function of the transmission distance $d$. State-discrimination receivers improve the 
SKR with respect to the heterodyne-based protocol in the regime $d \le 100$~km. In both pictures we set $\beta=0.95$.}\label{fig:02-KGR}
\end{figure}

In this section, we compare the results derived previously. In Figure~\ref{fig:02-KGR}(a) we plot the 
SKRs~(\ref{eq:Kopt}),~(\ref{eq:KPGM}) and~(\ref{eq:Khet}) as a function of the transmission distance $d$, expressed in km. The reconciliation efficiency is fixed to $\beta=0.95$ \cite{Lodewyck2005, Lodewyck2007, Denys2021}. It can be seen in the plot that both PGM and KOR beat the heterodyne-based protocol, that is $K_{\p} \ge K_{\HET}$, $\p=\PGM,\opt$. The improvement in the 
SKR is more relevant for metropolitan-network distances, in particular for $d\le 100$~km, whereas for larger ones both $K_{\opt}$ and $K_{\PGM}$ approach $K_{\HET}$ and achieve the same asymptotic scaling. To quantify this improvement we compute the ratio
\begin{eqnarray}
{\cal R}_{\p} = \frac{K_{\p}}{K_{\HET}} \, , \qquad (\p=\PGM,\opt) \, ,
\end{eqnarray}
reported in Figure~\ref{fig:02-KGR}(b). Both the ratios exhibit peaks for $d\le 40$~km and then decrease towards~$1$ in the long-distance regime, but the behaviour is rather different between the two cases. In fact, ${\cal R}_{\PGM}$ achieves a single maximum at $\approx 5$~km, increasing the 
SKR with respect to $K_{\HET}$ by more than $42 \%$, and then decays monotonously to~$1$. On the contrary, $K_{\opt}$ is not a monotonic function of the transmission distance and, in turn, the associated ratio exhibits two separated peaks. The KOR coincides with the PGM up to its first maximum, that is ${\cal R}_{\opt}={\cal R}_{\PGM}$ for $d \lesssim 7$~km, while for larger $d$ we have ${\cal R}_{\opt}\ge{\cal R}_{\PGM}$. Thereafter, ${\cal R}_{\opt}$ reaches a local minimum and then achieves a second maximum at $\approx 23$~km, with $\approx 47 \%$ increase in the 
SKR. Ultimately, the curve decreases to $1$, approaching the heterodyne-based protocol together with ${\cal R}_{\PGM}$.

\begin{figure}
\begin{center}
\includegraphics[width=\textwidth]{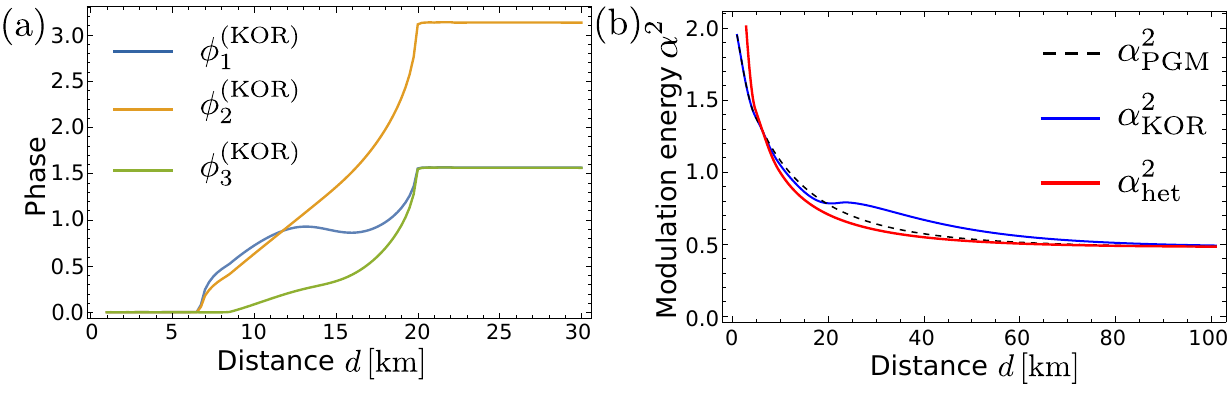}
\end{center}
\caption{(a) Plot of the optimized phases $\phi_j^{(\opt)}$, $j=1,\ldots,3$, as a function of the transmission distance $d$ in km. We recall that $\phi_0^{(\opt)}=0$. (b) Plot of the optimized modulation energies $\alpha^2_{\p}$, $\p=\PGM,\opt$, and $\alpha^2_{\HET}$, as a function of the transmission distance $d$.  In both pictures we set $\beta=0.95$.}\label{fig:03-OptPar}
\end{figure}

The behavior of $K_{\opt}$ is a consequence of the resulting optimized phases $\phi_j^{(\opt)}$, depicted in Figure~\ref{fig:03-OptPar}(a). We recall that $\phi_0^{(\opt)}=0$ by definition. For $d \lesssim 7$~km we have $\boldsymbol{\phi}_{\opt}={\bf 0}$ and the optimized receiver is identical to the PGM, whereas for larger distances the optimized phases are nonzero and ${\cal R}_{\opt}\ge{\cal R}_{\PGM}$. Interestingly, for $d \gtrsim 20$~km the optimized phase tuple becomes distance-independent and reads $\boldsymbol{\phi}_{\opt}=(0,\pi/2,\pi,\pi/2)$. This choice allows to reach the second maximum in Figure~\ref{fig:02-KGR}(b), after which the KOR approaches the heterodyne-based protocol.
For completeness, Figure~\ref{fig:03-OptPar}(b) reports also the optimized energies $\alpha^2_{\p}$, $\p=\PGM,\opt$, and $\alpha^2_{\HET}$. All curves converge to~$0.5$ average number of photons in the long-distance regime but, differently from the other cases, $\alpha^2_{\opt}$ shows the same non-monotonic trend of $K_{\opt}$.

\begin{figure}
\begin{center}
\includegraphics[width=0.6\textwidth]{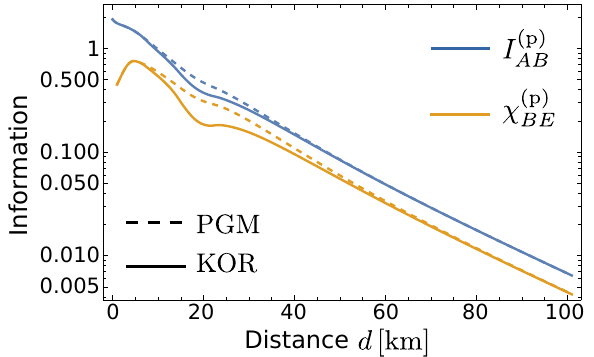}
\end{center}
\caption{Logarithmic plot of $I_{AB}^{(\p)}$ and $\chi_{BE}^{(\p)}$, $\p=\PGM,\opt$, as a function of the transmission distance $d$ in km. Both the quantities are computed with the optimized parameters $\alpha^2_{\p}$ and $\boldsymbol{\phi}_{\opt}$ (for the optimized receiver). We set $\beta=0.95$.}\label{fig:04-Info}
\end{figure}

The previous results prove non-Gaussian receivers as a potential tool for improving the key rate of the QPSK protocol, at least in the present composable security approach. Remarkably, they also highlight that the discrete-valued POVM minimizing the error probability, namely the PGM, does not coincide with the discrete-valued POVM maximizing the 
SKR, namely the KOR. 
The reason becomes evident when comparing separately the mutual and the Holevo information appearing in the 
SKR~(\ref{eq:KGR}). In Figure~\ref{fig:04-Info} we plot the quantities $I_{AB}^{(\p)}$ and $\chi_{BE}^{(\p)}$, $\p=\PGM,\opt$, computed with the same optimized energy and phases previously obtained and depicted in Figure~\ref{fig:03-OptPar}.
As we can see, in the metropolitan-network distance regime the optimized receiver is associated with a reduced mutual information with respect to the PGM but, at the same time, reducing the mutual information induces also a reduction of the Holevo information extractable by Eve, thus resulting in a higher 
SKR.
As a consequence, differently from the state-discrimination scenario, in CV-QKD there emerges a tradeoff between the goal of increasing the information accessible to Bob and the necessity of making the encoded symbols less ``distinguishable" to weaken Eve's attack.

In light of this, we may interpret the physical meaning of the optimized phases as follows.
For small transmission distances $\kappa d \ll 1$, Eve's intercepted signals are too weak to give her sufficient knowledge on which symbol was sent and the two different goals of reducing the error probability and maximizing the 
SKR are compatible, therefore the KOR coincides with the PGM. On the contrary, for larger $d$ the compatibility does not hold anymore, and Bob has to sacrifice part of his potential information and to reduce the mutual information shared with Alice to the detriment of the eavesdropper.

\begin{figure}
\begin{center}
\includegraphics[width=0.42\textwidth]{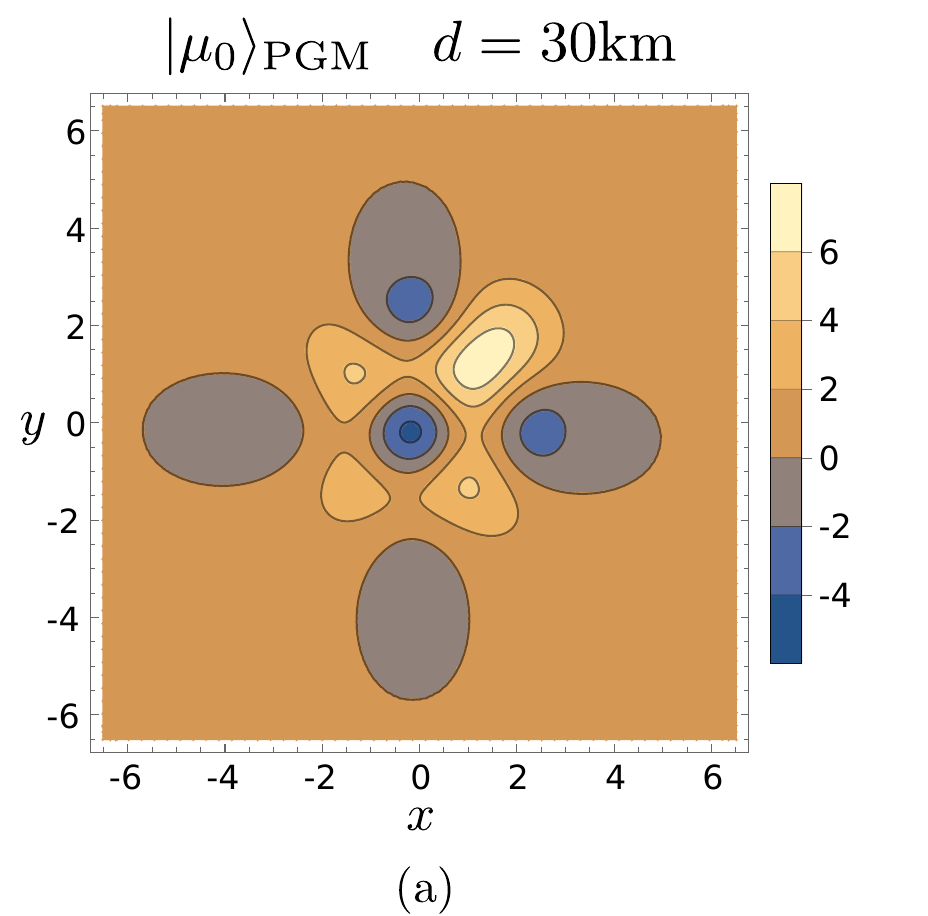} \quad
\includegraphics[width=0.42\textwidth]{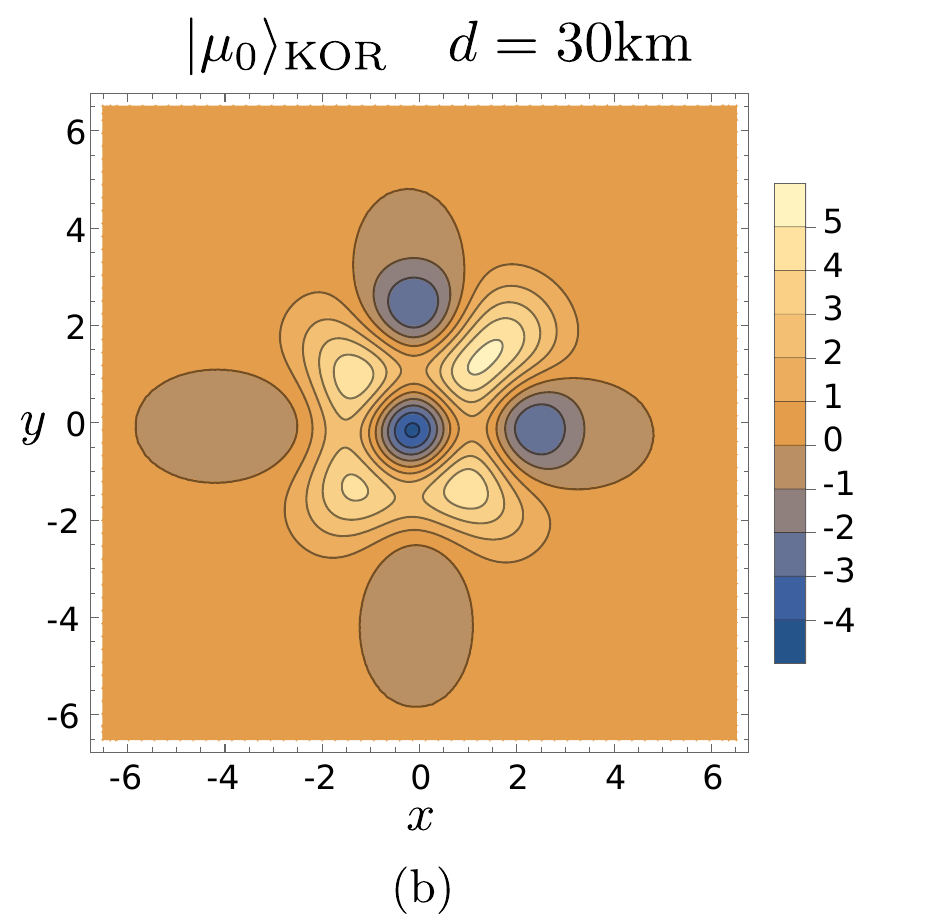} \\[2ex]
\includegraphics[width=0.42\textwidth]{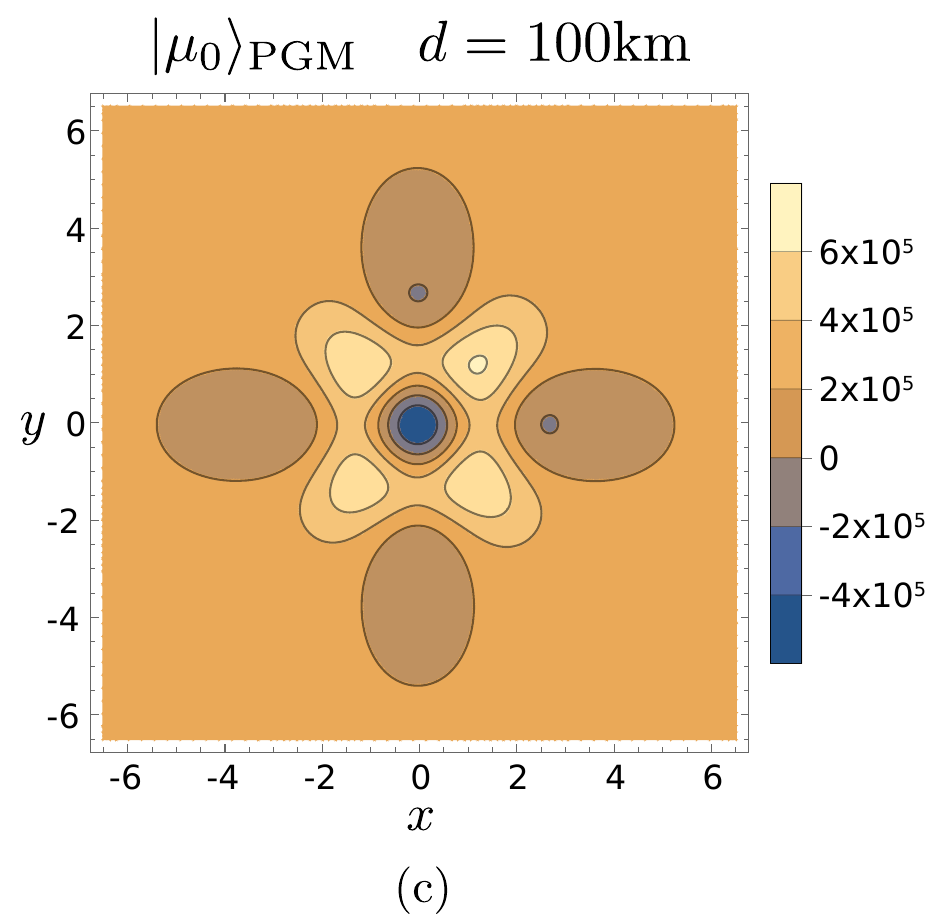} \quad
\includegraphics[width=0.42\textwidth]{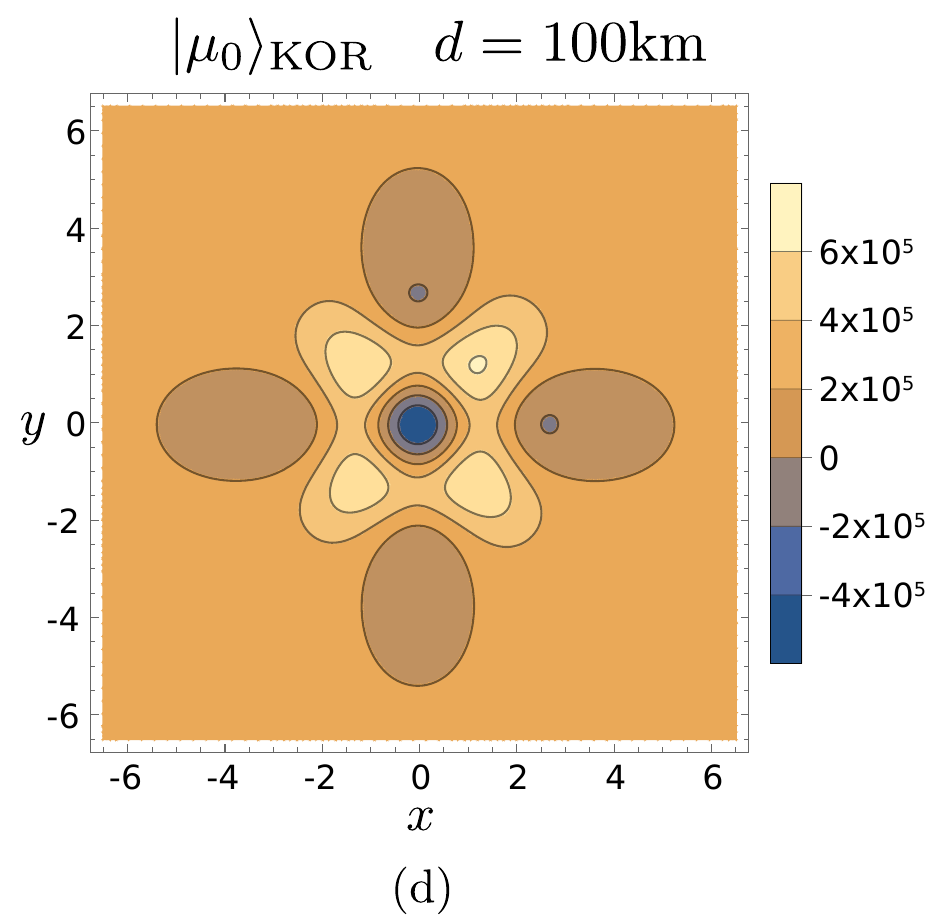}
\end{center}
\caption{Contour plot of the Wigner functions $W^{(\p)}(x,y)$ of the reference measurement vectors $|\mu_0\rangle_{\p}$, $\p=\PGM,\opt$, for either $d=30$ km (a-b) or $d=100$ km (c-d). We set $\alpha^2=1$ and $\boldsymbol{\phi}={\bf 0}$ and $\boldsymbol{\phi}=(0,\pi/2,\pi,\pi/2)$ for the PGM and the optimized receiver, respectively.}\label{fig:05-Wigner}
\end{figure}

The discussed tradeoff may be qualitatively appreciated by comparing the phase-space representations of the PGM and the KOR effects. More in detail, we consider the two reference measurement vectors $|\mu_0\rangle_{\p}$, $\p=\PGM,\opt$, computed from~(\ref{eq:mu0phi}) with the phases $\boldsymbol{\phi}={\bf 0}$ and $\boldsymbol{\phi}=(0,\pi/2,\pi,\pi/2)$, respectively, and compute the associated Wigner function
\begin{eqnarray}
W^{(\p)}(x,y) = \frac{2}{\pi} \sum_{n=0}^{\infty} \, (-1)^n \, \langle n | D\dag(\zeta) \, \rho_\p \, D(\zeta) | n\rangle \, , \quad (\p=\PGM,\opt) \, ,
\end{eqnarray}
where $\zeta= (x+ \rmi y)/2$ expressed in SNU, $\rho_\p= |\mu_0\rangle_{\p} {\, }_{\p}\langle \mu_0|$ and $D(\zeta)$ is the displacement operator \cite{Olivares2021, Serafini2017}. The contour plots of $W^{(\p)}(x,y)$ are depicted in Figure~\ref{fig:05-Wigner} for $\alpha^2=1$ and two different transmission distances $d=30$ km and $d=100$ km.
If $d=30$ km, that is for metropolitan-network distances, there is a qualitative difference between the two compared  cases, see Figures~\ref{fig:05-Wigner}(a) and~\ref{fig:05-Wigner}(b). Both Wigner functions exhibit four peaks, corresponding to the four transmitted states $|\alpha_k^{(t)}\rangle$. However, $W^{(\PGM)}(x,y)$ is well concentrated around state $|\alpha_0^{(t)}\rangle$, while $W^{(\opt)}(x,y)$ is more delocalized over the four states and the peaks are less distinguishable. This implies a reduced distinguishability of the states and, in turn, a reduced mutual information $I_{AB}^{(\opt)}$.
On the contrary, when the distance is larger, e.g. $d=100$ km, the transmitted states are weak coherent states with a greater overlap between one another. As a consequence, $W^{(\p)}(x,y)$ for respective receivers are equally delocalized over the four peaks and the differences between PGM and KOR become negligible; see Figures~\ref{fig:05-Wigner}(c) and~\ref{fig:05-Wigner}(d). In turn, the associated 
SKRs converge to the same value, corresponding also to the rate of the heterodyne-based protocol, as depicted in Figure~\ref{fig:02-KGR}.
Furthermore, in all cases we observe a Wigner-negativity, proving both $|\mu_0\rangle_{\p}$ to be non-classical (as well as non-Gaussian) states at all distances \cite{Olivares2021,Serafini2017,Genoni2013}.

\section{Employing feasible receivers}\label{sec:Izumi}
\begin{figure}
\begin{center}
\includegraphics[width=0.65\textwidth]{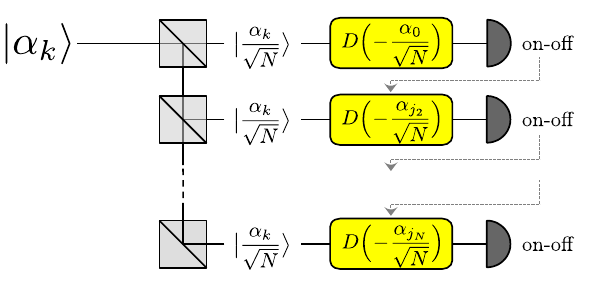}
\end{center}
\caption{Scheme of the displacement feed-forward receiver proposed in \cite{Izumi2012}. The incoming signal $|\alpha_k\rangle$ is split into $N$ copies. Each copy $m=1,\ldots, N$ undergoes a conditional displacement $D(-\alpha_{j_{m}}/\sqrt{N})$ followed by on-off detection. For the first copy we have $j_1=0$. For the others, the outcome of the $(m-1)$-th detection sets out the displacement amplitude $j_{m}$ to be implemented on the following copy.}\label{fig:06-DispRec}
\end{figure}

Even though both PGM and KOR discussed in the previous sections have shown interesting potentialities for CV-QKD, from a practical point of view there is no clear idea on their experimental implementation. 
Differently from the case of binary discrimination where the PGM is implemented via the Dolinar receiver \cite{Dolinar1973, Assalini2011}, in the presence of QSPK designing a feasible optimum receiver is an open problem. In  fact, the Dolinar receiver has been generalized, obtaining the so-called Bondurant receiver \cite{Bondurant1993}, which, unfortunately, is not optimum. As a consequence, it is not known how to reach the minimum error probability with optical feedback and linear optics.
Nevertheless, at the state of the art many feasible quantum receivers are based on displacement operations and photon counting \cite{Kennedy1973, Izumi2012, Becerra2013, DiMario2018-1,  DiMario2018-2, Izumi2020, Notarnicola2023}.
Therefore, it is worth of interest to investigate also the performance of these receivers for CV-QKD. Here, in particular, we focus on the proposal of the displacement feed-forward receiver (DFFRx) presented in \cite{Izumi2012} and depicted in Figure~\ref{fig:06-DispRec}. Its functioning will be presented in detail in the following subsection.

\subsection{Description of the displacement feed-forward receiver}
The 
DFFRx, presented schematically in Figure~\ref{fig:06-DispRec}, aims at QPSK discrimination of coherent states in the form $|\alpha_k\rangle = |\alpha \, \rme^{\rmi \pi (2k+1)/M} \rangle$, $k=0,\ldots, M-1$, with $M=4$. It is based on the slicing property of coherent states: indeed, thanks to an array of splitters, the incoming signal $|\alpha_k\rangle$ is split into $N \ge M-1$ identical copies with reduced amplitude $|\alpha_k/\sqrt{N}\rangle$. Then, each $m$-th copy, $m=1,\ldots, N$, undergoes a conditional displacement operation followed by an on-off detection which returns a click-no click result. The first copy is displaced by $D(-\alpha_{j_1}/\sqrt{N})$, with $j_1=0$, being mapped into the coherent state $|(\alpha_k-\alpha_{j_1})/\sqrt{N}\rangle$. In turn, if $k=0$ the incoming signal is displaced into the vacuum and the subsequent on-off detector will not click, whereas if $k\neq 0$ the detector is more likely to click with a probability $1-p_k$, where
\begin{eqnarray}\label{eq:pk}
p_0&=1 \,,  \nonumber \\
p_1&= p_3= \rme^{-2\alpha^2/N} \,,  \nonumber \\
p_2&= \rme^{-4\alpha^2/N} \, .
\end{eqnarray}
According to the result of the first detection, we decide what would be the value of the amplitude of the displacement $D(-\alpha_{j_2}/\sqrt{N})$ applied to the second copy: if an "off" result is registered, that is the detector does not click, we set $j_2=j_1=0$; otherwise we discard hypothesis "$k=0$", set $j_2=j_1+1$ and probe the final hypothesis from the remaining set $k=1,2,3$. We proceed iteratively in this way until the last copy is processed, following the feed-forward rule: if the $(m-1)$-th detection gives outcome ``off" we displace the $m$-th copy by $D(-\alpha_{j_{m}}/\sqrt{N})$ with $j_{m}=j_{m-1}$, if an ``on" is obtained we set $j_{m}=j_{m-1}+1$, discard all states $j\le j_{m-1}$ and restrict the decision to the states $j_m,\ldots, M-1$. The outcome of the last detection determines the final decision. If an ``off" is retrieved, we decide the state $j=j_m$ has been sent, otherwise we perform a random decision among the remaining states.

The conditional probabilities of inferring the state $j=0,\ldots,M-1$ if state $k=0,\ldots,M-1$ was sent read:
\numparts
\begin{eqnarray}
p_{B|\alpha_k}^{(N)}(0) =& p_0^N \, , \\[2ex]
p_{B|\alpha_k}^{(N)}(1) =& \sum_{t=0}^{N-2} p_k^t \, (1-p_k) \, p_{(k-1) \bmod M}^{N-1-t} 
\, + \, \frac{p_k^{N-1}(1-p_k)}{3} \, ,\\[2ex]
p_{B|\alpha_k}^{(N)}(2) =& \sum_{t=0}^{N-3} \, \sum_{s=0}^{N-3-t} p_k^t \, (1-p_k) \, p_{(k-1) \bmod M}^s \, (1-p_{(k-1) \bmod M}) \times \, \nonumber \\[1ex]
&\, p_{(k-2) \bmod M}^{N-2-t-s} \, + \, \sum_{t=0}^{N-2} p_k^t \, (1-p_k) \, \frac{p_{(k-1) \bmod M}^{N-2-t}(1-p_{(k-1) \bmod M})}{2} \nonumber \\[1ex]
&\, + \, \frac{p_k^{N-1}(1-p_k)}{3} \, ,\\[2ex]
p_{B|\alpha_k}^{(N)}(3) =& \sum_{t=0}^{N-3} \, \sum_{s=0}^{N-3-t}  \,\, \sum_{u=0}^{N-3-t-s}  p_k^t \, (1-p_k) \, p_{(k-1) \bmod M}^s \times \, \nonumber \\[1ex]
&\, (1-p_{(k-1) \bmod M}) \, p_{(k-2) \bmod M}^{u} \, (1-p_{(k-2) \bmod M}) \, p_{(k-3) \bmod M}^{N-3-t-s-u} \nonumber \\[1ex]
& \, + \, \sum_{t=0}^{N-2} p_k^t \, (1-p_k) \, \frac{p_{(k-1) \bmod M}^{N-2-t}(1-p_{(k-1) \bmod M})}{2} \nonumber \\[1ex]
&\, + \, \frac{p_k^{N-1}(1-p_k)}{3} \, .
\end{eqnarray}
\endnumparts
The overall Bob's probability can then be written as:
\begin{eqnarray}
p_B^{(N)}(j) =\frac{1}{M} \sum_{k=0}^{M-1} p_{B|\alpha_k}^{(N)}(j) \, , \quad (j=0,\ldots, M-1) \, .
\end{eqnarray}

\begin{figure}
\begin{center}
\includegraphics[width=0.55\textwidth]{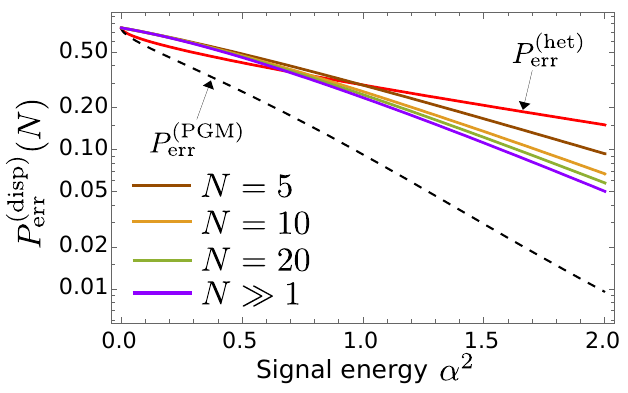}
\end{center}
\caption{Plot of the decision error probability $P_{\rm err}^{(\disp)}(N)$ as a function of the signal energy $\alpha^2$ for different $N$, compared to both the SQL and the minimum error probability achieved by the PGM. The displacement feed-forward receiver beats the SQL only in the regime $\alpha^2 \gg 1$ and, for large $N$, scales as $P_{\rm err}^{(\disp)}(N) \approx \alpha^2 \rme^{-2\alpha^2}$, whilst the minimum error probability is $P_{\rm err}^{(\PGM)} \approx \rme^{-2\alpha^2}$.}\label{fig:07-ErrorP}
\end{figure}

In the context of quantum-state discrimination, the 
associated decision error probability reads:
\begin{eqnarray}
P_{\rm err}^{(\disp)}(N)=1-\frac{1}{M}\sum_{k=0}^{M-1} p_{B|\alpha_k}^{(N)}(k)\, , 
\end{eqnarray}
depicted in Figure~\ref{fig:07-ErrorP} as a function of $\alpha^2$ for different $N$.
As emerges from the plot, the present displacement receiver outperforms the SNL achieved with heterodyne detection, namely $P_{\rm err}^{(\HET)}=1-[1+{\rm erf}(\alpha/\sqrt{2})]^2/4$, only in the high-energy regime $\alpha^2 \gg 1$.
We also note that, in the limit $N\gg1$, we have $P_{\rm err}^{(\disp)}(N) = \rme^{-2\alpha^2}(\alpha^2+3/4)$, and the DFFRx reaches the Bondurant receiver \cite{Bondurant1993}.
Nevertheless, it does not provide a near-optimum receiver. In fact, the minimum error probability, associated with the PGM, reads 
\begin{eqnarray}
P_{\rm err}^{(\PGM)}= 1- \frac{1}{M^2} \left(\sum_{k=0}^{M-1} \sqrt{\lambda_k}\right)^2 \, ,
\end{eqnarray}
with $\lambda_{0(1)}= 2 \rme^{-\alpha^2}[\cosh(\alpha^2)\pm \cos(\alpha^2)]$ and $\lambda_{2(3)}= 2 \rme^{-\alpha^2}[\sinh(\alpha^2)\pm \sin(\alpha^2)]$, which for $\alpha^2\gg 1$ scales only exponentially, $P_{\rm err}^{\rm (PGM)} \approx \rme^{-2\alpha^2}$.

When Bob employs the DFFRx in the protocol of Figure~\ref{fig:01-Protocol}, the previous conditional probabilities shall be modified accordingly by substituting $\alpha^2 \rightarrow T \alpha^2$ in the $p_k$ in~(\ref{eq:pk}), as Bob receives only the transmitted fraction of Alice's signals.

\subsection{Calculation of the 
SKR}
\begin{figure}
\begin{center}
\includegraphics[width=\textwidth]{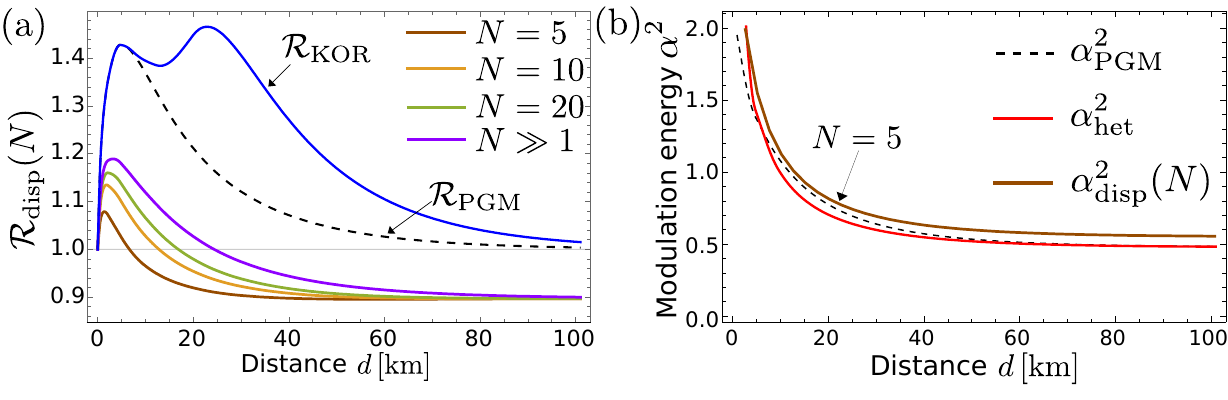}
\end{center}
\caption{(a) Plot of the ratio ${\cal R}_\disp(N)$ as a function of the transmission distance $d$ in km. Differently from both PGM and KOR, discussed in Section~\ref{sec:Results}, the 
DFFRx improves the 
SKR with respect to the heterodyne-based protocol only up to a maximum distance $d_{\rm max}(N)$, increasing with the number of copies $N$.
(b) Plot of the optimized modulation energies $\alpha^2_{\disp}(N)$, $\alpha^2_{\PGM}$, and $\alpha^2_{\HET}$, as a function of the transmission distance $d$.  In both pictures we set $\beta=0.95$. }\label{fig:08-Ratio}
\end{figure}

We now compute the 
SKR $K_\disp(N;\alpha^2)$ associated with the 
DFFRx by exploiting Equations~(\ref{eq:KGR}),~(\ref{eq:IAB}) and~(\ref{eq:chiBE}), provided the substitutions $p_{B|\alpha_k}^{(\boldsymbol{\phi})}\rightarrow p_{B|\alpha_k}^{(N)}$ and $p_{B}^{(\boldsymbol{\phi})}\rightarrow p_{B}^{(N)}$, and optimize over the modulation energy, obtaining:
\begin{eqnarray}
K_\disp(N) = \max_{\alpha^2} \,  K_\disp(N; \alpha^2) \, .
\end{eqnarray}
Moreover, we also compute the ratio with respect to the heterodyne-based protocol, namely,
\begin{eqnarray}
{\cal R}_\disp(N) = \frac{K_\disp(N)}{K_\HET} \, ,
\end{eqnarray}
reported in Figure~\ref{fig:08-Ratio}(a) for different number of copies $N$.
Unlike the PGM and the KOR, the 
DFFRx outperforms the heterodyne-based protocol only up to a maximum transmission distance $d_{\rm max}(N)$ whose value increases with $N$. Afterwards, $K_\disp(N) \le K_{\HET}$ and, in turn, ${\cal R}_\disp(N)$ saturates to an asymptotic value $\le 1$. The best performance is achieved in the limit of infinite copies, $N\gg 1$, where the receiver approximates the Bondurant receiver, obtaining a maximum increase in the 
SKR of about $\lesssim 20 \%$ and $d_{\rm max}(N) \lesssim 25$ km.

This behaviour is a direct consequence of the optimized modulation $\alpha^2_{\disp}(N)$, reported in Figure~\ref{fig:08-Ratio}(b). In fact, $\alpha^2_{\disp}(N)$ is a decreasing function of $d$, which in the long-distance regime, $\kappa d \gg 1$, reaches an asymptotic value $\gtrsim 0.5$. Numerical calculations also show this asymptote to be independent of the number of copies $N$. 
In these conditions, Bob receives a signal with $T \alpha^2_{\disp}(N) \ll 0.5$ mean photons, for which the DFFRx does not beat the SQL, as depicted in Figure~\ref{fig:07-ErrorP}. In turn, even the SKR of the CV-QKD protocol is lower than the corresponding heterodyne protocol.
On the other hand, for $\kappa d \ll 1$, the optimized modulation is of few photons, the DFFRx outperforms the SQL and we observe an increase also in the SKR.

In conclusion, despite its feasibility, the present displacement feed-forward scheme is not optimal for CV-QKD, just as it is not optimal for coherent state discrimination. Nevertheless, it still provides an improvement of the resulting key rate in the short-distance regime, being a candidate for experimental realizations of the present protocol.

\section{Conclusions and outlooks}\label{sec:Conc}
In this paper we have investigated the role of state-discrimination receivers for CV-QKD. In particular, we have addressed a QPSK protocol based on a pure-loss wiretap channel assumption, and proposed an optimized non-Gaussian state-discrimination receiver, the KOR, to be employed by Bob in place of the commonly exploited Gaussian measurements. We have compared the performance of both the KOR and PGM with respect to the heterodyne-based protocol and obtained an increase in the 
SKR for metropolitan-network distances. Moreover, we have also discussed the role of feasible schemes, such as the displacement feed-forward receiver, obtaining an increased 
SKR in the short-distance regime up to a maximum transmission distance.

The results obtained in this paper suggest suitable non-Gaussian receivers as a resource to increase the key rate in the CV-QKD framework.
Nevertheless, they only provide a first step towards the analysis of CV-QKD with non-Gaussian measurements and leave many points as open problems.
At first, the extension of the present analysis to the more realistic case of a thermal-loss channel remains a challenging task. In fact, in the presence of thermal mixed states, designing the receiver achieving the minimum error probability is non-trivial. The general structure of quantum receivers derived in Section~\ref{sec:QDT} does not hold anymore, as the $M$ mixed states now span the whole infinite dimensional Hilbert space. Moreover, from the perspective of quantum communications, the optimum receiver achieving the minimum error probability can only be obtained numerically via linear convex semidefinite programming \cite{Cariolaro2015}.
There exists an extension of the PGM method, where the Gram matrix is obtained via the ``factor decomposition" of the quantum states \cite{Cariolaro2015, Assalini2010, Eldar2004, Cariolaro2010}. However, the resulting POVM is not optimal anymore, even in the presence of GUS, and only provides a non-tight upper bound of the minimum error probability. As a consequence, the search of the key-rate optimized receiver could only be obtained via a brute-force functional optimization over all possible POVMs, being a nonlinear and non-convex problem.

Secondly, the sketch of an unconditional security proof may be designed, identifying which is the optimal Eve's attack. To do so, we should optimize over all the possible attacks compatible with Alice and Bob's statistics, retrieving the DW bound \cite{Denys2021, Devetak2005} by extending the methods of \cite{Lin2019}. Indeed, the question whether or not protocols employing non-Gaussian measurement guarantee higher security than Gaussian ones is an interesting open problem.

Finally, we should investigate the scalability of the present scheme with discrete-modulation formats of higher order, like PSK schemes with $M\ge 4$ encoded states or quadrature-amplitude-modulation (QAM) constellations, in which the GUS is not satisfied anymore \cite{Cariolaro2015, Notarnicola2022}.

\ack
We thank C.~Marquardt for insightful discussions.
This project has received funding from the European Union’s Horizon Europe research and innovation programme under the project ``Quantum Security Networks Partnership" (QSNP, grant agreement No~101114043).

\appendix
\section{Derivation of the constraint on matrix $A$}\label{app:AdagA}

As discussed in Section~\ref{sec:QDT}, the optimal receiver for pure-state discrimination is a 1-rank POVM $\{\Pi_j\}_j$, $\Pi_j=|\mu_j\rangle\langle \mu_j|$, $j=0,\ldots, M-1$, where $|\mu_j\rangle = \sum_k a_{kj} |\gamma_k\rangle$, $a_{kj} \in \mathbb{C}$ and
\begin{eqnarray}\label{eq:NormPOVM}
{\cal N} \equiv \sum_{j=0}^{M-1} |\mu_j\rangle\langle \mu_j| = \mathbb{P}_{\cal S} \, .
\end{eqnarray}
The previous condition may be turned into the constraint~(\ref{eq:ConstA}) as follows. Equation~(\ref{eq:NormPOVM}) is an equality of quantum operators, therefore it shall be
\begin{eqnarray}
{\cal N} |\psi\rangle = |\psi\rangle \, \qquad \forall \, |\psi\rangle \in {\cal S} \, ,
\end{eqnarray}
where $|\psi\rangle = \sum_s b_s |\gamma_s\rangle$, $b_s \in \mathbb{C}$. Accordingly, the following equations hold:
\begin{eqnarray}
\Bigg[\sum_j \Big( \sum_{k,l} a_{kj} a_{lj}^*  |\gamma_k\rangle \langle \gamma_l | \Big) \Bigg] \sum_s b_s |\gamma_s\rangle = \sum_t b_t |\gamma_t\rangle \, , \\[1ex]
\sum_k \Big(\sum_{j,l,s} a_{kj} a_{jl}\dag G_{ls} b_s \Big) |\gamma_k\rangle = \sum_t b_t |\gamma_t\rangle \, ,
\end{eqnarray}
where $G_{ls}= \langle \gamma_l|\gamma_s\rangle$. In the matrix notation we have $A A\dag G  \, {\bf b} = {\bf b}$ to be satisfied for all ${\bf b} = (b_0,\ldots, b_{M-1})$. In turn,
\begin{eqnarray}
A A\dag G = \Id_M \, ,
\end{eqnarray}
$\Id_M$ being the $M \times M$ identity matrix and, ultimately, $A A\dag = G^{-1}$.

\section{Computation of the von Neumann entropy of Eve's states}\label{app:Entropies}
To perform the 
SKR analysis outlined in Section~\ref{sec:POVM} we need to compute the von Neumann entropies of states~(\ref{eq:rhoE}) and~(\ref{eq:rhoEcondB}). Both of them are in the following form:
\begin{eqnarray}\label{eq:state}
\rho = \sum_{k=0}^{M-1} c_k \, |\alpha_k^{(r)}\rangle \langle \alpha_k^{(r)}| \, ,
\end{eqnarray}
for some coefficients $c_k \in\mathbb{C}$. To compute the associated entropy we need to diagonalize state~(\ref{eq:state}). If $|\psi\rangle$ is the eigenvector of $\rho$ associated with eigenvalue $\lambda$, we have $|\psi\rangle = \sum_m b_{m} |\alpha_m^{(r)}\rangle$ and the following equations hold:
\begin{eqnarray}
\rho |\psi\rangle = \lambda |\psi\rangle  \\[1ex]
\Bigg( \sum_k c_k |\alpha_k^{(r)}\rangle \langle \alpha_k^{(r)}| \Bigg) \sum_m b_{m} |\alpha_m^{(r)}\rangle =  \lambda  \sum_s b_{s} |\alpha_s^{(r)}\rangle  \\[1ex]
\sum_k  c_k \Bigg( \sum_m  G_{km} b_{m}   \Bigg) |\alpha_k^{(r)}\rangle =  \lambda  \sum_s b_{s} |\alpha_s^{(r)}\rangle \, ,
\end{eqnarray}
where $G_{km}= \langle \alpha_k^{(r)}|\alpha_m^{(r)}\rangle$.

As a consequence, we have a set of equations:
\begin{eqnarray}
\lambda \, b_{k} = c_k \Bigg( \sum_{m=0}^{M-1}  G_{km} b_{m}  \Bigg) \, , \quad (k=0,\ldots, M-1) \, ,
\end{eqnarray}
or, equivalently,
\begin{eqnarray}
\Bigg(\frac{\lambda}{c_k}-1 \Bigg) b_{k} -  \sum_{m\neq k} G_{km} b_{m} = 0  \, .
\end{eqnarray}
This defines the homogeneous linear system $M {\bf b}= 0$, where ${\bf b}=(b_0,\ldots, b_{M-1})$ and
\begin{eqnarray}
M= 
\left(
\begin{array}{cccc} 
\frac{\lambda}{c_0}-1 & -G_{01} & -G_{02} & -G_{03} \\
-G_{10} &\frac{\lambda}{c_1}-1 &  -G_{12} & -G_{13} \\
-G_{20} & -G_{21} & \frac{\lambda}{c_2}-1 & -G_{23} \\
-G_{30} & -G_{31} & -G_{32} & \frac{\lambda}{c_3}-1 \\
\end{array}
\right) \, .
\end{eqnarray}
The equation $M {\bf b}= 0$ always admits a trivial solution ${\bf b}= 0$, therefore to obtain a nonzero eigenvector we shall impose the condition $\det M = 0$. This provides us with the four eigenvalues $\{ \lambda_j \}_j$ and the corresponding von Neumann entropy $S[\rho] =~-\sum_j \lambda_j \log_2 \lambda_j$.

For state $\rho_E$ in~(\ref{eq:rhoE}), for which $c_k= M^{-1}$, the equation $\det M = 0$ may be solved analytically, leading to:
\begin{eqnarray}
\lambda_{0(1)}&= \frac{\rme^{-(1-T) \alpha^2}}{2} \Bigg\{\cosh \Big[(1-T) \alpha^2\Big] \pm \cos\Big[(1-T) \alpha^2\Big]\Bigg\} \, , \nonumber \\[1.5ex]
\lambda_{2(3)}&= \frac{\rme^{-(1-T) \alpha^2}}{2} \Bigg\{ \sinh \Big[(1-T) \alpha^2\Big] \pm \Big|\sin\Big[(1-T) \alpha^2\Big]\Big|\Bigg\} \, .
\end{eqnarray}

\section*{References} 

\bibliographystyle{iopart-num}

\providecommand{\newblock}{}

\end{document}